\documentclass[11pt,dvips]{article} 
\usepackage{amssymb,url,hyperref}
\title{Photon number variance in isolated cavities} 
\author{
Fabrice \textsc{Philippe}
\thanks{Fabrice \textsc{Philippe}, 
Universit\'e Paul Val\'ery, Route de Mende,  34199 MONTPELLIER Cedex 5 - FRANCE}
\ and Jacques \textsc{Arnaud}
\thanks{Jacques \textsc{Arnaud}, Mas Liron, 30440 
SAINT-MARTIAL - FRANCE} \\ 
\small\textit {MIAp, Universit\'e Paul Val\'ery, Montpellier, France}\\ 
\small\textit {and  LIRMM, Montpellier, France}\\
\small{e-mail: Fabrice.Philippe@univ-montp3.fr } }

\begin{document}
\maketitle
\vspace{1cm}
\begin{abstract}
We consider a strictly isolated single-mode optical cavity resonating at angular
frequency $\omega$ containing atoms whose one-electron level energies are
supposed to be: $\hbar\omega$, $2\hbar\omega$\ldots $B\hbar\omega$, and $m$
photons.  If initially the atoms are in their highest energy state and $m=0$, we
find that at equilibrium: variance$(m)/$mean$(m)=(B+1)/6$, indicating that the
internal field statistics is sub-Poissonian if the number of atomic levels $B$
does not exceed 4.  Remarkably, this result does not depend on the number of
atoms, nor on the number of electrons that each atom incorporates.  Our result
has application to the statistics of the light emitted by pulsed lasers and
nuclear magnetic resonance.  On the mathematical side, the result is based on
the restricted partitions of integers.
\end{abstract}

\section{Introduction}

We consider a single-mode optical cavity containing identical atoms.  The number
of photons in the cavity, denoted by $m$, may be measured at any time $t$: It
suffices in principle to quickly remove the atoms at that time and introduce an
ideal light detector.  The photo-count tells us how many photons were present in
the cavity at the time considered.  By performing measurements on a large number
of similarly prepared cavities, the probability $P(m,t)$ that some $m$-value be
found at time $t$ is evaluated.  We are particularly interested in the so-called
Fano factor, defined for example in \cite{kn:y91}: $F(t)\equiv$
variance$(m)/$mean$(m)$.  In the course of time, the system eventually reaches a
state of equilibrium, in which case $P(m,t)$ and $F(t)$ are time-independent.

If the system is in a state of \emph{thermal} equilibrium, the photon statistics
is, as is well known, that of Bose-Einstein.  In that case, the Fano factor is
equal to mean$(m)+1$.  This situation occurs when the cavity may exchange energy
with a thermal bath (canonical ensemble).  This would also be the case for the
strictly isolated cavities (micro-canonical ensemble) considered in the present
paper if the response of the atoms to the field were \emph{linear}.  But in
general the atomic response is nonlinear and the photon distribution does not
follow the Bose-Einstein statistics.

This may be seen from a simple example.  Consider an isolated single-mode cavity
resonating at angular frequency $\omega$ and containing two identical (but
distinguishable) resonant 2-level atoms.  The atomic ground-state energy is
taken as equal to 0 and the upper level energy as equal to $\hbar \omega=1$, for
brevity.  If, initially, the two atoms are in their upper state and the cavity
does not contain any photon, i.e., $m$(0)=0, the total matter+field energy
$U=2$.  Part of the atomic energy gets converted into photons in the course of
time.  The fundamental law of Statistical Mechanics tells us that, once a state
of equilibrium has been reached, all the microstates of the isolated system
presently considered are equally likely to occur.  The complete list of
microstates (first atom state, second atom state, number of photons) reads:
(1,1,0), (0,1,1), (1,0,1) and (0,0,2).  It follows from this list that the
probabilities of having 0, 1 and 2 photons are proportional to 1, 2 and 1,
respectively.  This is obviously a non-thermal distribution.  In that example,
the Fano factor defined above is $F=1/2$.  But the mean value of $m$ is equal to
1, so that the Bose-Einstein distribution would give instead $F=2$.  Sets of
microstates may be obtained similarly for total energy values $U=0,1,2,3\ldots$. 
When the cavity is in contact with a bath at temperature $T$, the probability
that the total energy be $U$ is given by the Boltzmann probability law
$\exp(-U/T)$.  The end result for the photon statistics is of course
Bose-Einstein.  Only isolated cavities with a particular value of the total
energy $U$ are presently considered.  It is straightforward to generalize the
two 2-level atoms result to any number $M$ of distinguishable 2-level atoms.  We
find that $F$ remains equal to 1/2.  This is a special case ($B=2$) of the
general result to be derived in the present paper.  The Bose-Einstein statistics
would, in that case, give: $F=1+M/2$.

The Fano factor has been evaluated in many papers dealing with laser light, for
example \cite{kn:y91, kn:htw89}.  In lasers, the atoms are driven to their
excited state by a pump, and the cavity suffers from some optical loss, perhaps
as a result of light transmission through partially reflecting mirrors.  It
could be thought at first that the Fano factor of isolated cavities obtains from
the laser theory result by letting the pumping rate $J$ as well as the optical
loss $\alpha$ go to zero.  This is not the case, however, because in the laser
system the total (atom + field) energy $U$ in the cavity may drift slowly, no
matter how small $J$ and $\alpha$ are.  It follows that the variance of $m$
deduced from laser theories in that limit does not coincide with the present
Statistical Mechanical result, applicable to strictly isolated systems, even if
the average value of $U$ is the same in both cases.  This point has been
discussed in detail in \cite{arnaud}.  Further conceptual details can be found
in \cite{kn:acp00}.  The present theory is nevertheless of practical
significance.  It is applicable to \emph{pulsed} rather than continuous
electromagnetic generators, as we later discuss.

To summarize, if initially $m=0$ and the piece of matter is in its highest
energy state, $m$ at some later time represents the energy subtracted from the
piece of matter.  The probability that some $m$ value occurs is proportional to
the matter statistical weight (number of distinguishable configurations) $W_{m}$
according to the equal-weight principle of Statistical Mechanics.  We therefore
need only consider the statistical weight of the atomic collection.  We will
consider first the case of a single atom with level energies 1,2\ldots,$B$, and
subsequently any number, $M$, of $B$-level atoms.  Finally, we consider for the
sake of illustration a radio-frequency cavity containing $M$ bismuth nuclei
(spin 9/2) immersed in a magnetic field.  Our general result gives in that case
a Fano factor equal to 11/6.

Our simple and general expression for the Fano factor derives from a property of
the number of restricted partitions of integers.

\section{One $B$-level atom}
Consider an atom whose one-electron level energies are $1,2,\ldots,B$, with
$N\leq B$ single-spin electrons.  According to the Pauli exclusion principle
each level may be occupied by only 0 or 1 electron.  The atom energy is greatest
when the $N$ electrons occupy the upper states.  For some subtracted energy $m$
the statistical weight $W_{m}$ is the number $p(N,m)$ of partitions of $m$ into
at most $N$ parts, none of them exceeding $B-N$.  This conclusion is reached by
shifting electrons downward beginning with the bottom one until the specified
energy $m$ is subtracted.  Let us recall that a partition of $m$ is a
non-increasing sequence of positive integers summing up to $m$.  By convention
the number of partitions of 0 is 1.  It is known \cite{kn:a76} that
\begin{equation}\label{eq:1}
	g(q)\equiv \sum_{m\geq 0}p(N,m)q^m=
	\prod_{i=1}^{B-N}\frac{1-q^{N+i}}{1-q^i}.
\end{equation}

If the moments of $m$ with respect to the statistical weight $W_{m}$ are defined
as
\begin{equation}\label{eq:2}
	\overline{m^n}=\frac{\sum_{m=0,1,2\ldots}m^np(N,m)}
	{\sum_{m=0,1,2\ldots}p(N,m)},	
\end{equation}
we have, as it is well known \cite{kn:p91},  
\begin{eqnarray}
	\mathrm{mean}(m) & = & \overline{m}=h'(1),
	\label{eq:3a}  \\
	\mathrm{variance}(m) & = & \overline{m^2}-\overline{m}^2=h''(1)+h'(1),
	\label{eq:3b}
\end{eqnarray}
where $h(q)=\ln[g(q)]$ and primes (double primes) denote first (second)
derivatives with respect to the argument.

Since 
\begin{equation}
	\ln(\frac{1-q^{N+i}}{1-q^i})=
\ln(1+\frac{N}{i})+\frac{N}{2}(q-1)+\frac{N}{24}(N+2i-6)(q-1)^2+O((q-1)^3),
	\label{eq:4}
\end{equation}
we obtain after summation over $i$ from 1 to $B-N$ 
\begin{eqnarray}
	\mathrm{mean}(m) & = & h'(1)=\frac{1}{2}N(B-N),
	\label{eq:5a}  \\
	\mathrm{variance}(m) & = &h''(1)+h'(1)=\frac{1}{12}N(B-N)(B+1),
	\label{eq:5b}
\end{eqnarray}
\begin{equation}
	\frac{\mathrm{variance}(m)}{\mathrm{mean}(m)}=\frac{B+1}{6}.
	\label{eq:6}
\end{equation}
For example, if $B=4, N=2$, a direct examination shows that $p(2,m)=1$ if
$m=0,1,3,4$ and $p(2,2)=2$.  Therefore, $\mathrm{mean}(m)=2$ and
$\mathrm{variance}(m)=10/6$ in agreement with (\ref{eq:6}).  If the equilibrium
field is allowed to radiate into free-space the statistics of the emitted
photons is sub-Poissonian (variance less than the mean) when $B< 5$ and
super-Poissonian (variance exceeding than the mean) when $B> 5$.  The result in
(\ref{eq:6}) was given in \cite{kn:acp00} without a proof.

\section{Any number of $B$-level atoms}

Let now the cavity contain a collection of $M$ atoms labeled by $k=1,2\ldots,M$,
with $N^{(k)}\le B$ electrons in atom $k$.  These atoms are supposed to be
distinguishable and to be coupled to one another only through energy exchanges
with the cavity field.

The photon number $m$ represents the energy subtracted from the atomic
collection.  The atomic statistical weight $W_{m}$ is the sum, for all values of
$m_{1},m_{2},\ldots$ summing up to $m$, of the products of the individual
statistical weights defined above:
\begin{equation}
	W_{m}=\sum_{m_{1}+m_{2}+\ldots+m_{M}=m}
	p(N^{(1)},m_{1})p(N^{(2)},m_{2})\ldots p(N^{(M)},m_{M}).
	\label{eq:}
\end{equation}
The moments of $m$ may therefore be calculated as 
\begin{equation}\label{eq:9}
	\overline{m^n}=\frac{\sum(m_{1}+m_{2}+\ldots+m_{M})^n
	p(N^{(1)},m_{1})p(N^{(2)},m_{2})\ldots p(N^{(M)},m_{M})} 
	{\sum p(N^{(1)},m_{1})p(N^{(2)},m_{2})\ldots p(N^{(M)},m_{M})},
\end{equation}
where the summation is over all non-negative values of $m_{1}, m_{2}, \cdots,
m_{M}$.  It follows that the mean value of $m$ is the sum of the individual
atoms means, and that the variance of $m$ is the sum of the individual atoms
variances.

For the $B$-level atoms considered here, the result (\ref{eq:5a}) gives
\begin{equation}\label{eq:8}
\mathrm{mean}(m)=\sum_{k=1}^M \mathrm {mean}^{(k)}
=\sum_{k=1}^M\frac{N^{(k)}(B-N^{(k)})}{2}, 
\end{equation}
and from (\ref{eq:5b})
\begin{equation}\label{eq:10}
\mathrm{variance}(m)=\sum_{k=1}^M\mathrm{variance}^{(k)}
=\sum_{k=1}^M\frac{N^{(k)}(B-N^{(k)})(B+1)}{12}.
\end{equation}
Therefore, the simple result in (\ref{eq:6}) holds for any collection of
$B$-level atoms.

\section{Application to nuclear magnetic resonance}

As is well known, a spin 1/2 charged particle such as an electron immersed in a
magnetic field behaves in the same manner as a (one-electron) 2-level atom. 
This analogy generalizes to spin-$s$ particles.  We may therefore consider, as
an example of application of the previous expressions, an electro-magnetic
cavity containing $M$ identical spin-$s$ charged particles.  These particles may
be distinguished from one another by their locations.  If these particles are
submitted to a magnetic field of appropriate strength, and in appropriate energy
units, the energy levels are $-s,-s+1,\cdots,s$.  In a cold environment only the
lowest energy levels are populated.  But it suffices to apply the so-called
$\pi$-radio-frequency pulse to get the highest levels populated.  Our previous
result: variance$(m)/$mean$(m)=(B+1)/6=(s+1)/3$ applies once a state of
equilibrium between the particles and the field has been reached.  It is here
supposed that the nuclei natural relaxation time is much longer than the time
required for the particle-field equilibrium to be attained.  If the field is
allowed to radiate into free space, the emitted electro-magnetic pulse is
Poissonian (variance$(m)=$mean$(m)$) for spin 2 particles.

Bismuth nuclei were found by Black and Goudsmit in 1927 to have a maximum spin
$s=9/2$ \cite{kn:lt91}.  It follows that in the presence of a magnetic field
these nuclei exhibit $B=2s+1=10$ evenly-spaced energy levels.  When located in a
radio-frequency cavity (whose resonating frequency should be in the 100 MHz
range for usual magnetic-field strengths), the equilibrium Fano factor reads
according to our theory $F=(B+1)/6=11/6$.  Because of their small energy,
radio-frequency photons may be counted only at very low temperatures.  It is
also at such low temperatures that long nuclei relaxation times may occur.

{\small \section*{Acknowledgement} 
The authors wish to express their thanks to L. Chusseau and D. Scalbert for
critical readings of the manuscript.

}
\end{document}